\documentclass[a4paper,11pt]{amsart}
\usepackage{graphicx}
\begin{document}
\hyphenation{gra-vi-ta-tio-nal re-la-ti-vi-ty Gaus-sian
re-fe-ren-ce re-la-ti-ve gra-vi-ta-tion Schwarz-schild
ac-cor-dingly gra-vi-ta-tio-nal-ly re-la-ti-vi-stic pro-du-cing
de-ri-va-ti-ve ge-ne-ral ex-pli-citly des-cri-bed ma-the-ma-ti-cal
de-si-gnan-do-si coe-ren-za pro-blem gra-vi-ta-ting geo-de-sic
per-ga-mon cos-mo-lo-gi-cal gra-vity cor-res-pon-ding
de-fi-ni-tion phy-si-ka-li-schen ma-the-ma-ti-sches ge-ra-de
Sze-keres con-si-de-red}
\title[Highest-Energy Cosmic Rays and Hilbertian Repulsive Effect]
{{\bf Highest-Energy Cosmic Rays\\and Hilbertian Repulsive
Effect}}

\author[Angelo Loinger]{Angelo Loinger}
\address{A.L. -- Dipartimento di Fisica, Universit\`a di Milano, Via
Celoria, 16 - 20133 Milano (Italy)}
\author[Tiziana Marsico]{Tiziana Marsico}
\address{T.M. -- Liceo Classico ``G. Berchet'', Via della Commenda, 26 - 20122 Milano (Italy)}
\email{angelo.loinger@mi.infn.it} \email{martiz64@libero.it}

\vskip0.50cm

\begin{abstract}
We point out that an important portion of the high energy of the
cosmic rays from extragalactic sources can be attributed to a
Hilbertian repulsive effect, which is a consequence of Einstein
equations \emph{without} cosmological term.
\end{abstract}

\maketitle


relativity.

\normalsize

\vskip1.20cm \noindent Here is the abstract of a recent article on
\emph{Correlation of the Highest-Energy Cosmic rays with Nearby
Extragalactic Objects} \cite{1}: ``Using data collected at the
Pierre Auger Observatory during the past 3.7 years, we
demonstrated a correlation between the arrival directions of
cosmic rays with energy above $6\times10^{19}$ electron volts and
the positions of active galactic nuclei (AGN) lying within $\sim
75$ megaparsecs. We rejected the hypothesis of an isotropic
distribution of these cosmic rays with at least a $99\%$
confidence level from a prescribed a priori test. The correlation
we observed is compatible with the hypothesis that the
highest-energy particles originate from nearby extragalactic
sources whose flux has not been substantially reduced by
interaction with the cosmic background radiation
$[$Greisen-Zatsepin-Kuzmin effect$]$. AGN or objects having a
similar spatial distribution are possible sources.'' And in the
summary the authors write that ``AGN have long been considered as
likely sources of cosmic rays. Our data suggest that they remain
the prime candidates.''

\par First of all, we remember that in astrophysics we know
several boosting mechanisms for the charged particles, in
particular the \emph{magnetic reconnections} -- \emph{i.e.},
decaying magnetic fluxes --, and the \emph{Fermi accelerations}:
the motions in ionized gas clouds originate extended magnetic
fields of low field strength in the interstellar spaces;  charged
particles passing through these clouds gain, in statistical
average, more energy than they lose. Now, the cosmic rays with the
highest energy are prevalently protons and heavier nuclei.

\par AGN activity requires nuclear burning and magnetic
reconnections, as Kundt emphasizes \cite{2}. According to a
widespread belief, the core of the AGN would be a supermassive BH.
However, we have proved (\cite{3}, \cite{4}) that \emph{if we take
properly into account a hydrodynamical pressure} $p=p(t)$, any
gravitational collapse of a spherosymmetrical material ends in a
body with the \textbf{\emph{finite}} radius $(9/8)2m$, if $m\equiv
GM/c^{2}$ and $M$ is the gravitating mass. In papers \cite{4} and
\cite{5} we have illustrated the existence of a ``Hilbert effect''
\cite{6}, according to which in particular instances and in
particular regions the Einsteinian gravity -- \emph{without} the
intervention of a cosmological term -- exerts a
\textbf{\emph{repulsive}} action.

\par The differential equation of the \emph{radial} geodesics of
test-particles and light-rays in the Schwarzschild field of a
gravitating centre has the following first integral (\cite{4},
\cite{5}, \cite{6}):

\begin{equation} \label{eq:one}
\left(\frac{\textrm{d}r}{c \, \textrm{d}t}\right)^{2} =
\left(\frac{r-2m}{r}\right)^{2} \left[1+ A
\left(\frac{r-2m}{r}\right) \right] \quad,
\end{equation}

where $r$ is the \emph{standard} radial coordinate. $A$ is a
constant which is zero for light-rays and negative for
test-particles. For $(2/3)\leq|A|\leq1$, we have two regions, the
region in which the acceleration is negative (attractive gravity)
and the region in which the acceleration is positive
(\emph{repulsive} gravity), according to the following values of
the velocity:

\begin{equation} \label{eq:two}
\left|\frac{\textrm{d}r}{c \, \textrm{d}t}\right| <
\frac{1}{\sqrt3} \,  \frac{r-2m}{r} \quad, \quad
\textrm{(attraction)} \quad;
\end{equation}

\begin{equation} \label{eq:three}
\left|\frac{\textrm{d}r}{c \, \textrm{d}t}\right| >
\frac{1}{\sqrt3}  \, \frac{r-2m}{r} \quad, \quad
\textrm{(repulsion)} \quad.
\end{equation}

In reality, the case $|A|=2/3$ is a limiting case: the maximal
value of $(\textrm{d}r/c \, \textrm{d}t)^{2}$ is $(1/3)$, at
$r=\infty$, and the entire geodesic lies in a \emph{repulsive}
region. All the radial geodesics for which $\varepsilon \leq
|A|\leq2/3$ -- where $\varepsilon>0$ is an arbitrarily small
quantity -- belong entirely to a \emph{repulsive} region. And we
see that if $|A|=\varepsilon$, at $r=\infty$ we have
$|\,\textrm{d}r/\textrm{d}t|=\sqrt{1-\varepsilon}\cdot c$ ! The
following diagrams, where $x:=r/2m$, and $y(x):=(\textrm{d}r/c \,
\textrm{d}t)^{2}$, are of a great evidence.

\par For $x=9/8$, we have $y(9/8)=(1/9)^{2}[1-|A|(1/9)]$; if
$|A|=\varepsilon$, we have
$[y(9/8)]^{1/2}=(1/9)\sqrt{1-\varepsilon/9}$, \emph{i.e.}
$|\,\textrm{d}r/\textrm{d}t|=(1/9)\sqrt{1-\varepsilon/9}\, c$.
This is the initial velocity $v_{0}$ of a test-particle
$\textbf{\emph{T}}$ which arrives at $r=\infty$ with a velocity
$\sqrt{1-\varepsilon}\,c$. If for $\textbf{\emph{T}}$ we take,
\emph{e.g.} a proton of a cosmic ray, we must answer the question:
what mechanism did give it the remarkable velocity $v_{0}$ ? We
think that the \emph{magnetic reconnections} of the AGN generating
the cosmic rays yield a boosting mechanism that is quite
appropriate also from the \emph{quantitative} point of view
\cite{2}.

\par In conclusion, we set forth the assumption that an important
portion of the enormous energy of the cosmic rays from
extragalactic AGN can be attributed to a Hilbertian repulsive
effect \cite{7}. --

\vskip0.80cm
\par We are very grateful to our friend Dr. S. Antoci,
who sent us a copy of the paper by the \emph{Pierre Auger
Collaboration}.

\newpage
\begin{figure}[!hbp]
\begin{center}
\includegraphics[width=1.0\textwidth]{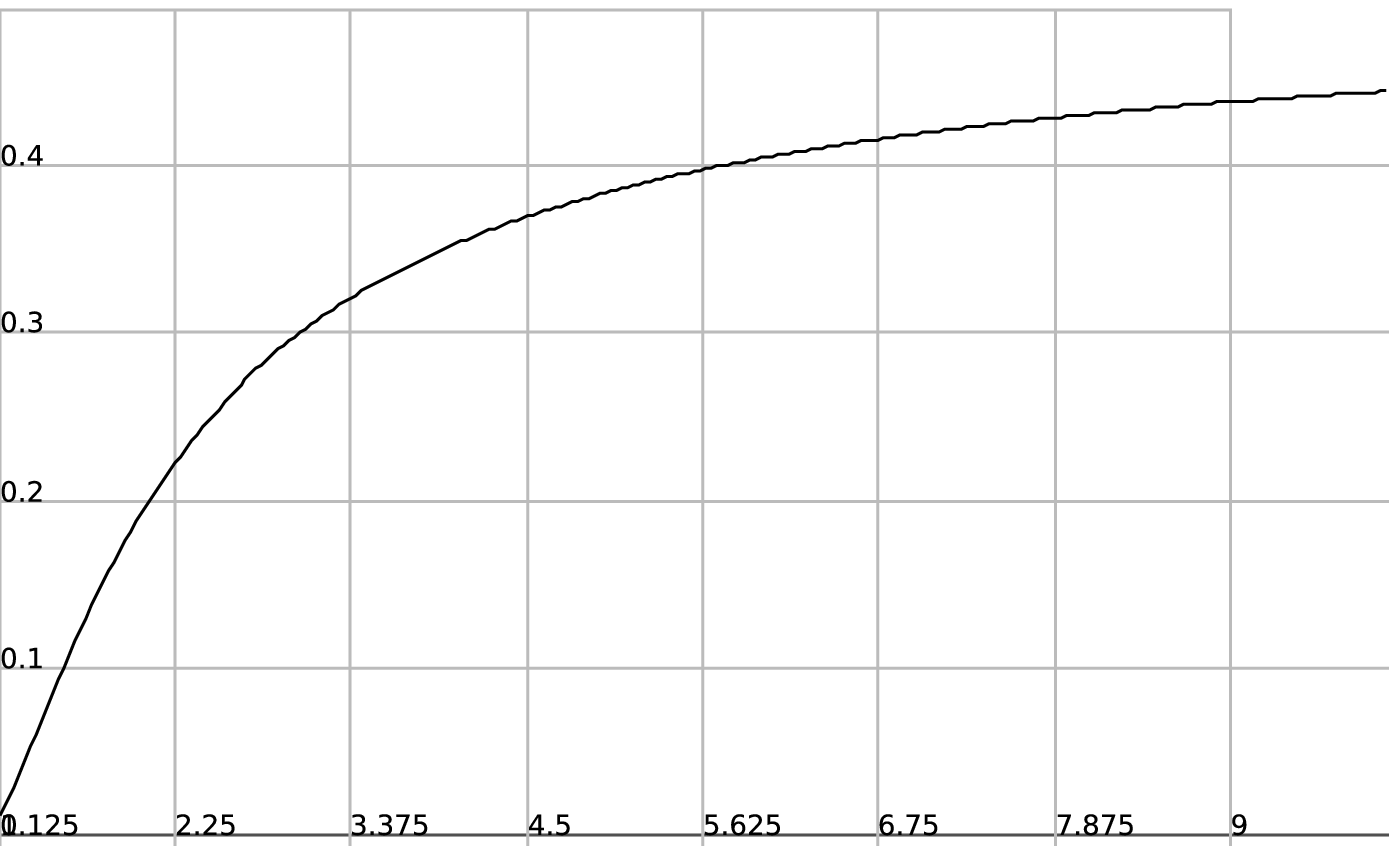}
\caption{\newline Diagram of $y(x)=[(x-1)/x]^{2}[1-0.5*(x-1)/x]$
for some values of $x$; $(9/8)\leq x <+\infty$;
$\max(+\infty;0.5)$; $[y(9/8)]^{1/2}=0.107981$.} \vskip1.00cm
\includegraphics[width=1.0\textwidth]{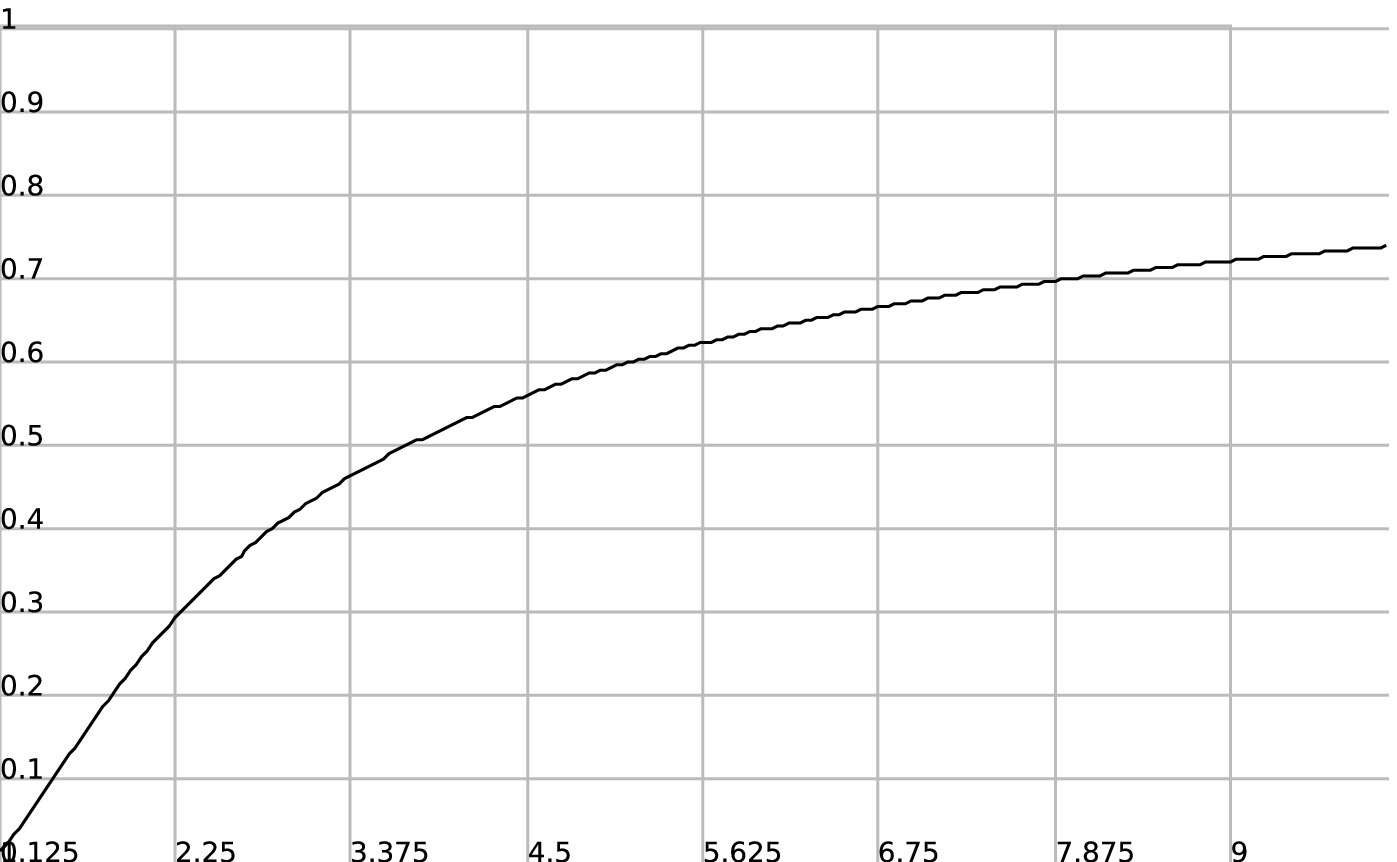}
\caption{\newline Diagram of
$y(x)=[(x-1)/x]^{2}[1-10^{-1}*(x-1)/x]$ for some values of $x$;
$(9/8)\leq x <+\infty$; $\max(+\infty;0.9)$;
$[y(9/8)]^{1/2}=0.110492$.}
\end{center}
\end{figure}

\newpage
\begin{figure}[!hbp]
\begin{center}
\includegraphics[width=1.0\textwidth]{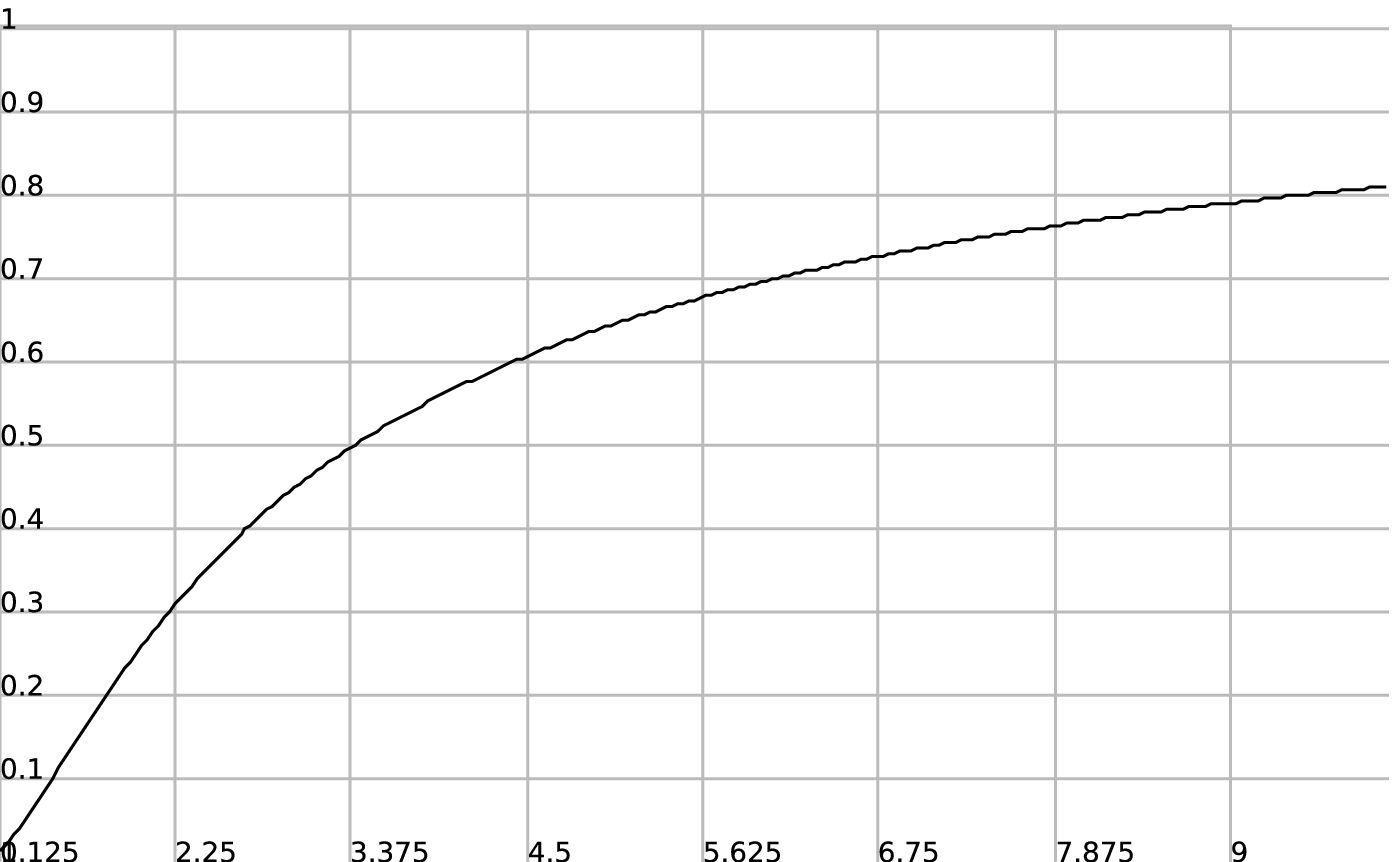}
\caption{\newline Diagram of
$y(x)=[(x-1)/x]^{2}[1-10^{-3}*(x-1)/x]$ for some values of $x$;
$(9/8)\leq x <+\infty$; $\max(+\infty;1-10^{-3})$;
$[y(9/8)]^{1/2}=0.111105$.} \vskip1.00cm
\includegraphics[width=1.0\textwidth]{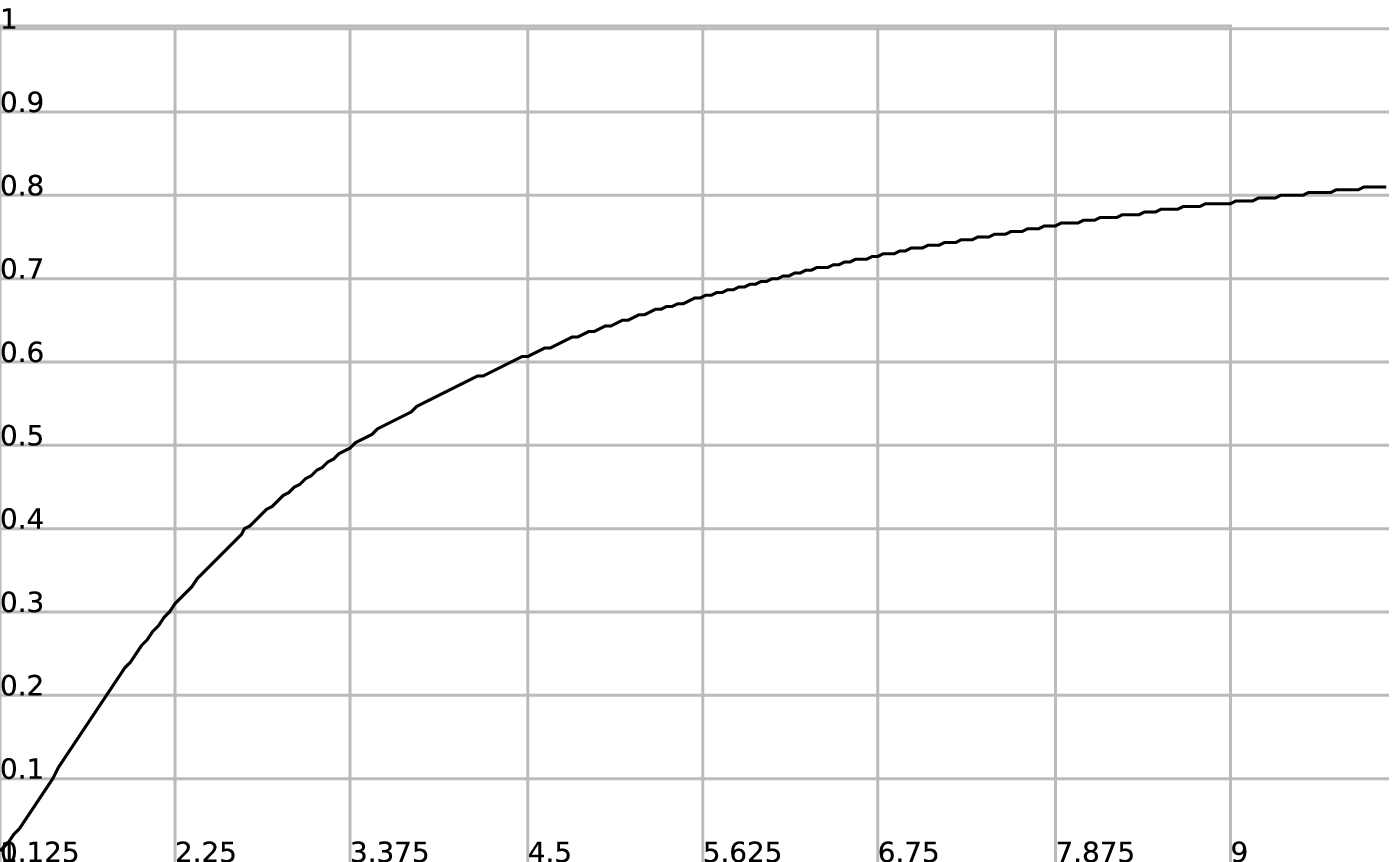}
\caption{\newline Diagram of
$y(x)=[(x-1)/x]^{2}[1-10^{-6}*(x-1)/x]$ for some values of $x$;
$(9/8)\leq x <+\infty$; $\max(+\infty;1-10^{-6})$;
$[y(9/8)]^{1/2}=0.111111$.}
\end{center}
\end{figure}

\end{document}